\documentclass[preprint,prc,aps,epsfig]{revtex4}

\usepackage{epsfig}
\def\beq{\begin{equation}}
\def\enq{\end{equation}}
\def\beqa{\begin{eqnarray}}
\def\enqa{\end{eqnarray}}

\def\GeV{\nobreak\,\mbox{GeV}}

\newcommand\tag{\hbox to hsize}

\begin{document}

\title{\sc Can the meson cloud explain the nucleon strangeness?}
\author{Fabiana Carvalho$^1$, Fernando S. Navarra$^2$ and Marina Nielsen$^2$}
\affiliation{$^1$ Instituto de F\'\i sica Te\'orica, Universidade
Estadual Paulista\\
Rua Pamplona 145, 01405-000 S\~ao Paulo, SP, Brazil\\
$^2$Instituto de F\'{\i}sica, Universidade de S\~ao Paulo, \\
C.P. 66318, 05315-970 S\~ao Paulo, SP, Brazil}

\begin{abstract}
We use a version of the meson  cloud model, including the kaon and the $K^*$
contributions, to estimate the electric and magnetic
strange form factors of the nucleon. We compare our results with the recent
measurements of the strange quark contribution to parity-violating asymmetries
in the forward G0 electron-proton scattering experiment. We conclude that it is 
very important to determine experimentally the electric and magnetic strange 
form factors, and not only the combination $G_E^s+\eta~G_M^s$, if one does 
really intend to understand the strangeness of the nucleon. 

\end{abstract}

\maketitle

As new experimental data appear, our picture of the nucleon evolves continually.
Our knowledge about the sea quarks in the nucleon has been changing 
dramatically and, in particular, our ideas about the strange sea quarks have 
been  
modified very rapidly. The famous EMC experiment \cite{emc} and other 
polarized DIS
experiments \cite{dis2} could be interpreted  as showing that the quarks carry 
only 
a small fraction of the total angular momentum of the proton. A further 
conclusion was 
that the strange sea quarks  in the proton are strongly polarized opposite to 
the 
polarization of the proton \cite{elliska}.  The recent results of the HERMES 
collaboration \cite{hermes} indicated that there is a SU(3) symmetry breaking 
in the 
nucleon sea. Most of these findings could be well understood with a meson 
cloud model (MCM) \cite{thomas1,thomas,thomas2,speth,na96}. 
In any version of the meson cloud model, the physical nucleon contains virtual 
meson-baryon components, 
which ``dress'' the bare nucleon. 
The meson cloud mechanism provides a natural explanation for 
symmetry breaking among  parton distributions \cite{ca00}. 
In \cite{thomas}, it has been shown that the inclusion  
of the meson cloud  significantly lowers the value of the total spin carried by
 quarks and antiquarks. In \cite{thomas}  the strange   
cloud was composed  by  $\Lambda K$ and $\Sigma K$ components in the Fock 
wavefunction of the proton and the 
authors obtained a very small polarization of the strange sea. 
Later on, in 
\cite{signal1,signal2},  it was shown that the higher mass components $\Lambda 
K^*$ and 
$\Sigma K^*$ could have  important effects on the strange sea. These components
 are 
kinematically suppressed but have large couplings  to the nucleon and may lead 
to a numerically 
significant contribution to some  observables. In particular, the states 
containing $K^*$ 
affect the quark-antiquark symmetry breaking  in the polarized strange sea. 
When only $K$ mesons
were considered it was observed that $x(\Delta s (x) -\Delta \overline{s} (x))>
 0$. When both 
contributions of $K$ and $K^*$ were included, as it was shown in \cite{signal2},
 $x(\Delta s (x) -\Delta \overline{s} (x))< 0$. 

Complementary to the high energy regime of \cite{emc,dis2,hermes} the nucleon 
strange sea can 
be probed in the low energy parity violating experiments carried out at TJNAF, 
where it is possible to
measure the strange electric and magnetic form factors of the nucleon. The first
 measurements of these quantities (and combinations of them) were performed by 
the SAMPLE 
\cite{sample} and HAPPEX \cite{happex1} collaborations. In  this low energy 
regime the strange 
component of the nucleon sea is expected to have a   nonperturbative origin. 
One of the 
possible nonperturbative mechanisms of strangeness production is given 
precisely by the meson 
cloud. Indeed, these data were studied in a number of approaches, including 
the MCM. 

Already in the first  kaon-cloud models the
nucleon strangeness distribution was  generated by fluctuations of the
``bare''  nucleon into kaon-hyperon intermediate states
which were described by the corresponding one loop Feynman graphs \cite{had}.
Since then, some concerns have been raised in the literature regarding the 
implementation of the loop model of the nucleon. In particular, it has been 
pointed 
out that truncations of the Fock space, which stop at the one-loop order, 
violate
unitarity \cite{ram97c}. While this is true in principle, the region where
rescattering should become important is above the production threshold,
which is at high momenta compared with those most relevant to the current 
process. Concerns have also been raised about the omission of
contributions from higher-lying intermediate states in the meson-hyperon 
fluctuations \cite{bar98,mel97}. While the effects of heavier hyperons,
such as the $\Sigma^*$, have been shown to be negligible \cite{mel97}, the
contribution of the $K^{\ast }-Y$ pairs were found to be large \cite{bar98}. 
Nevertheless,  the results of \cite{hil00} pointed to a ``slow 
convergence'' of the intermediate state sum.

The very recent results from G0 Collaboration at TJNAF \cite{g0}, provide 
information
on the nucleon strange vector form factors over the range of momentum transfers
$0.12\leq Q^2\leq 1.0~\GeV^2$. The data indicate non-trivial, $Q^2$ dependent,
strange quark distributions inside the nucleon, and present a challenge to
models of the nucleon structure.

In view of these new data we think that it is interesting to update our 
previous 
calculations  of the strange form factors with the  meson cloud model. One of 
the 
central issues will be the role played by the $K^*$ contribution. Our previous 
results indicated that the $K$ and $K^*$ loops lead to an opposite  $Q^2$ 
dependence of the combination $G_E^s(Q^2)+0.39 G_M^s(Q^2)$.
This conclusion is qualitatively consistent with 
the findings of \cite{signal2}, where an analogous  statement concerning the 
quantity $x(\Delta s (x) -\Delta \overline{s} (x))$ could be made.

In this brief report we will compute  the momentum dependence of the strange 
vector form factors in the loop model at momentum transfers $0\leq Q^{2} 
\leq3~ \GeV^2$, evaluated in ref.~\cite{hil00}, to compare with the results 
from G0 Coll. \cite{g0}. Although we believe that the results of this version
of the MCM
are more suitable for the momentum region $Q^2\leq1~\GeV^2$,
we extend our analysis up to $Q^2=3\GeV^2$ since new
experiments are being planned to cover this higher region of momentum transfer.

The nucleon matrix element of the strangeness current is parametrized by two
invariant amplitudes, the Dirac and Pauli strangeness form factors $%
F_{1,2}^{(s)}$:
\begin{equation}
\langle N(p^{\prime })|\bar{s}\gamma _{\mu }s|N(p)\rangle ={\bar{U}}%
(p^{\prime })\left[ F_{1}^{(s)}(Q^{2})\gamma _{\mu }+i{\frac{\sigma _{\mu
\nu }q^{\nu }}{2m_{N}}}F_{2}^{(s)}(Q^{2})\right] U(p)\ ,
\end{equation}
where $U(p)$ denotes the nucleon spinor and $F_{1}^{(s)}(0)=0$, due to the
absence of an overall strangeness charge of the nucleon.  The electric and
magnetic form factors are defined through
\beq
G_{E}^{(s)}(Q^{2})=F_{1}^{(s)}(Q^{2})-{Q^{2}\over4m_{N}^{2}}\,F_{2}^{(s)}
(Q^{2}),
\;\;\;\;\;\;G_{M}^{(s)}(Q^{2})=F_{1}^{(s)}(Q^{2})+F_{2}^{(s)}(Q^{2}).
\enq

We consider a hadronic one loop model containing $K$
and $K^{\ast }$ mesons as the dynamical framework for the calculation of
these form factors. This model is based on the meson--baryon effective
lagrangians 
\begin{eqnarray}
{\cal L}_{MB} &=&-g_{ps}\bar{B}i\gamma _{5}BK\ \ \ ,  \label{1aa} \\
{\cal L}_{VB} &=&-g_{v}\left[ \bar{B}\gamma _{\alpha }BV^{\alpha }-\frac{%
\kappa }{2m_{N}}\bar{B}\sigma _{\alpha \beta }B\partial ^{\alpha }V^{\beta }%
\right] \;,  \label{la}
\end{eqnarray}
where $B$ ($=N,\Lambda ,\Sigma $), $K$, and $V^{\alpha }$ are the baryon,
kaon, and $K^{\ast }$ vector meson fields, respectively, $m_{N}=939$ MeV is
the nucleon mass and $\kappa $ is the ratio of tensor to vector coupling, $%
\kappa =g_{t}/g_{v}$. In order to account for the finite extent of the above
vertices, the model includes form factors from the Bonn--J\"{u}lich $N-Y$
potential \cite{hol89} at the hadronic $KNY$ and $K^{\ast }NY$ ($Y=\Lambda
,\Sigma $) vertices, which have the monopole form 
\begin{equation}
F(k^{2})=\frac{m_M^{2}-\Lambda_M ^{2}}{k^{2}-\Lambda_M ^{2}}  \label{ff}
\end{equation}
with meson momenta $k$ and the physical meson masses $m_{K}=495$ MeV and $%
m_{K^{\ast }}=895$ MeV.
 
Since the non-locality of the meson-baryon form factors (\ref{ff}) gives
rise to vertex currents, gauge invariance was maintained in \cite{bar98} by
introducing the photon field via minimal substitution in the momentum
variable $k$ \cite{ohta}. The resulting
nonlocal seagull vertices are given explicitly in \cite{bar98,hil00}.

The diagonal couplings of $\bar{s}\gamma _{\mu }s$ to the strange mesons and
baryons in the intermediate states are straightforwardly determined by
current conservation, i.e. they are given by the net strangeness charge of
the corresponding hadron. The situation is more complex for the
non--diagonal (i.e. spin--flipping) coupling $F_{KK^{\ast }}^{(s)}(0)$ of
the strange current to $K$ and $K^{\ast }$, which is defined by the
transition matrix element

\begin{equation}
\langle K_{a}^{\ast }(k_{1},\varepsilon )|{\overline{s}}\gamma _{\mu
}s|K_{b}(k_{2})\rangle =\frac{F_{KK^{\ast }}^{(s)}(Q^{2})}{m_{K^{\ast }}}%
\,\epsilon _{\mu \nu \alpha \beta }\,k_{1}^{\nu }\,k_{2}^{\alpha
}\,\varepsilon ^{\ast \beta }\,\delta _{ab}\;.  \label{spinfl}
\end{equation}
This coupling was estimated in \cite{bar98} 
with the result $F_{KK^{\ast }}^{(s)}(0)=1.84$. The other couplings in the
effective Lagrangians are taken from the Nijmegen $NY$ potential 
\cite{nag77,rij98}:
 $g_{ps}/\sqrt{4\pi }=-4.005$, $g_{v}/\sqrt{4\pi }=-1.45$, $\kappa =2.43$, 
and 
we will consider the cutoff parameter values $\Lambda_K=0.9~\GeV$ and
 $0.9\GeV\leq\Lambda_{K^*}\leq1.1\GeV$ \cite{hai98}.
A smaller value for $g_{ps}$ was found in ref.~\cite{bra99}.

From the difference between the experimental asymmetry, measured by the
G0 Coll., and the ``no-vector-strange'' asymmetry, the combination
\beq
G_E^s(Q^2)+\eta(Q^2) G_M^s(Q^2),
\label{comb}
\enq
was obtained in ref.~\cite{g0}. In Eq.(\ref{comb})
$\eta(Q^2)=\tau G_M^p/\epsilon G_E^p$, with $\epsilon=(1+2(1+\tau)\tan^2
(\theta/2))^{-1}$, $\tau=Q^2/4m_N^2$ and $G_{E,M}^p$ being the electromagnetic
form factors of Kelly \cite{ke04}.

\begin{figure} \label{fig1}
\centerline{\psfig{figure=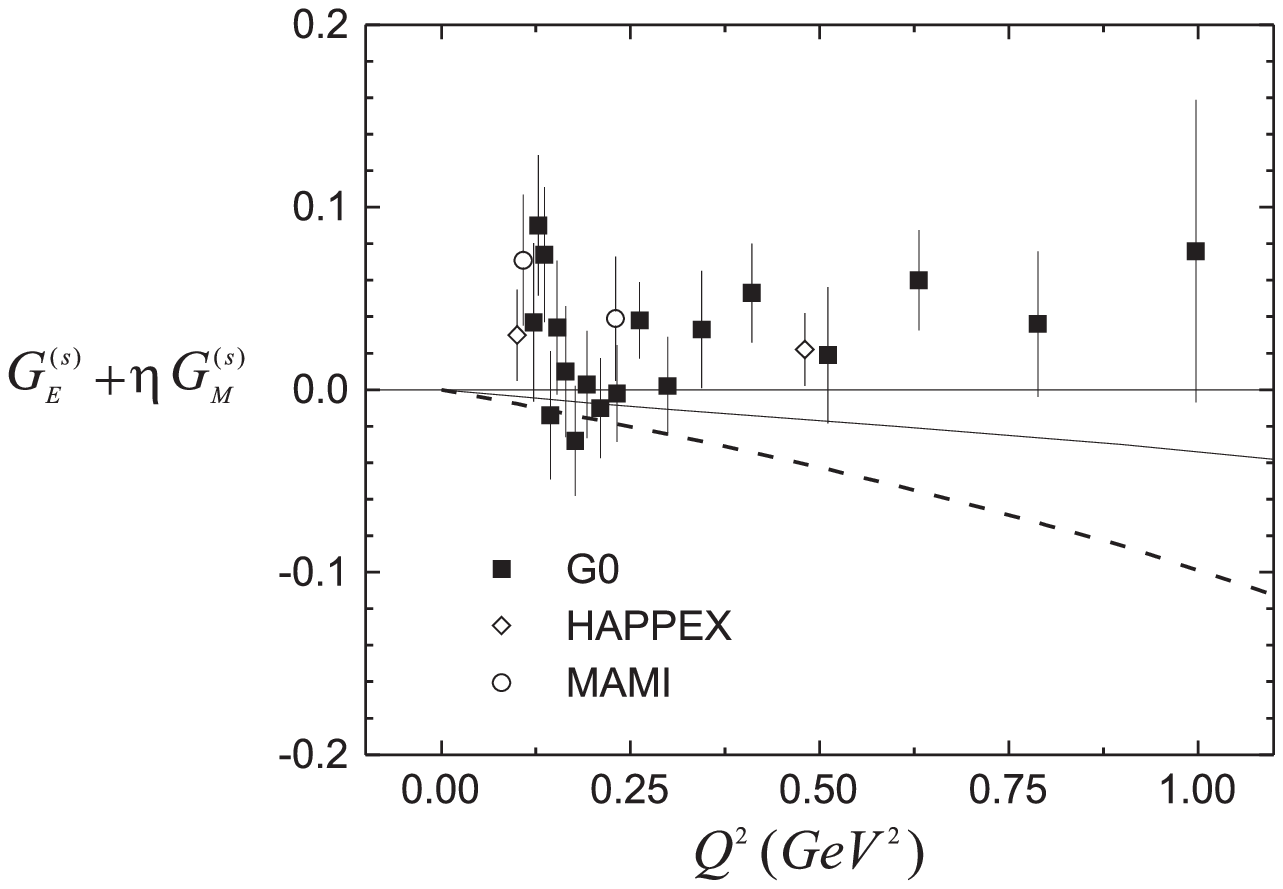,height=55mm,width=80mm,angle=0}
\psfig{figure=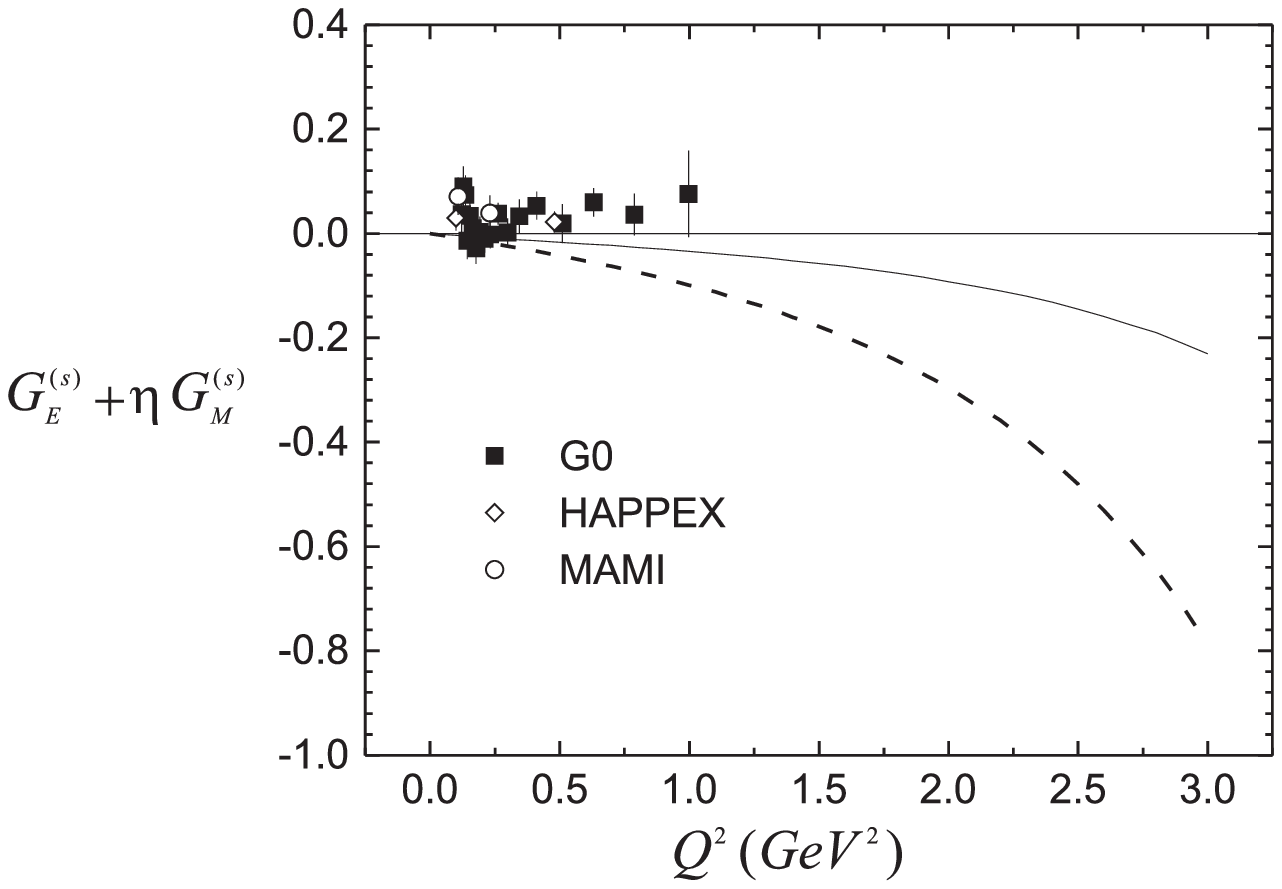,height=55mm,width=80mm,angle=0}}
\caption{The combination $G_E^s+\eta~ G_M^s$ as measured by the G0 Coll.
The solid and dashed lines show our  results with $\Lambda_K=\Lambda_{K^*}
=0.9\GeV$ and $\Lambda_K=0.9\GeV,~\Lambda_{K^*}=1.1\GeV$ respectively. In the 
left panel we show our results only up to $Q^2=1~\GeV^2$, to give a better 
view of the experimental data.} 
\end{figure}

In Fig. 1 we show the results of the loop model obtained by using 
$\Lambda_K=\Lambda_{K^*}=0.9~\GeV$ (solid line) and $\Lambda_K=0.9~\GeV,\;
\Lambda_{K^*}=1.1~\GeV$ (dashed line). We see that, although not completely
inconsistent with the G0 data (which seems to be consistent with zero), our 
results are negative and drecreasing with $Q^2$.
The cutoff value $\Lambda_{K^*} =0.9$ GeV, is very close
to the $K^{\ast }$ mass. As a consequence, for this cutoff value the
contributions from the $K^{\ast }$ and the $K/K^{\ast }$ transition are
completely negligible relative to the kaon contribution. Using a bigger
value for $\Lambda_{K^*}$ makes the agreement with the G0 data worse,
as can be seen by the dashed line in Fig.~1.
In Fig.1 we also show the new HAPPEX \cite{happex} and MAMI \cite{A4} data , 
which are in a very good agreement with the G0 data.

The HAPPEX Collaboration \cite{happex} has also estimated the values of
the electric and magnetic strange form factors at $Q^2\sim0.1~\GeV^2$. They
found: $G_E^s=-0.01\pm0.03$ and $G_M^s=0.55\pm0.28$. Using the SAMPLE result
for $G_M^s$: $G_M^s(Q^2=0.1)=0.37\pm0.22$ \cite{sample2}, the A4 
Collaboration at MAMI \cite{A4} got
$G_E^s(Q^2=0.108)=0.032\pm0.051$ and $G_E^s(Q^2=0.23)=0.061
\pm0.035$.  Our calculation gives
for these form factors at $Q^2\sim0.1~\GeV^2$ (with $\Lambda_K=\Lambda_{K^*}
=0.9\GeV$):  $G_E^s=0.0053$ and $G_M^s=-0.11$. Therefore, while we get 
$G_E^s$ compatible with data, our $G_M^s$ is  negative for the choice of 
parameters given above. A negative 
value for $G_M^s$ was also obtained in a recent lattice calulation \cite{de05}.
\begin{figure}[h] \label{fig2}
\centerline{\psfig{figure=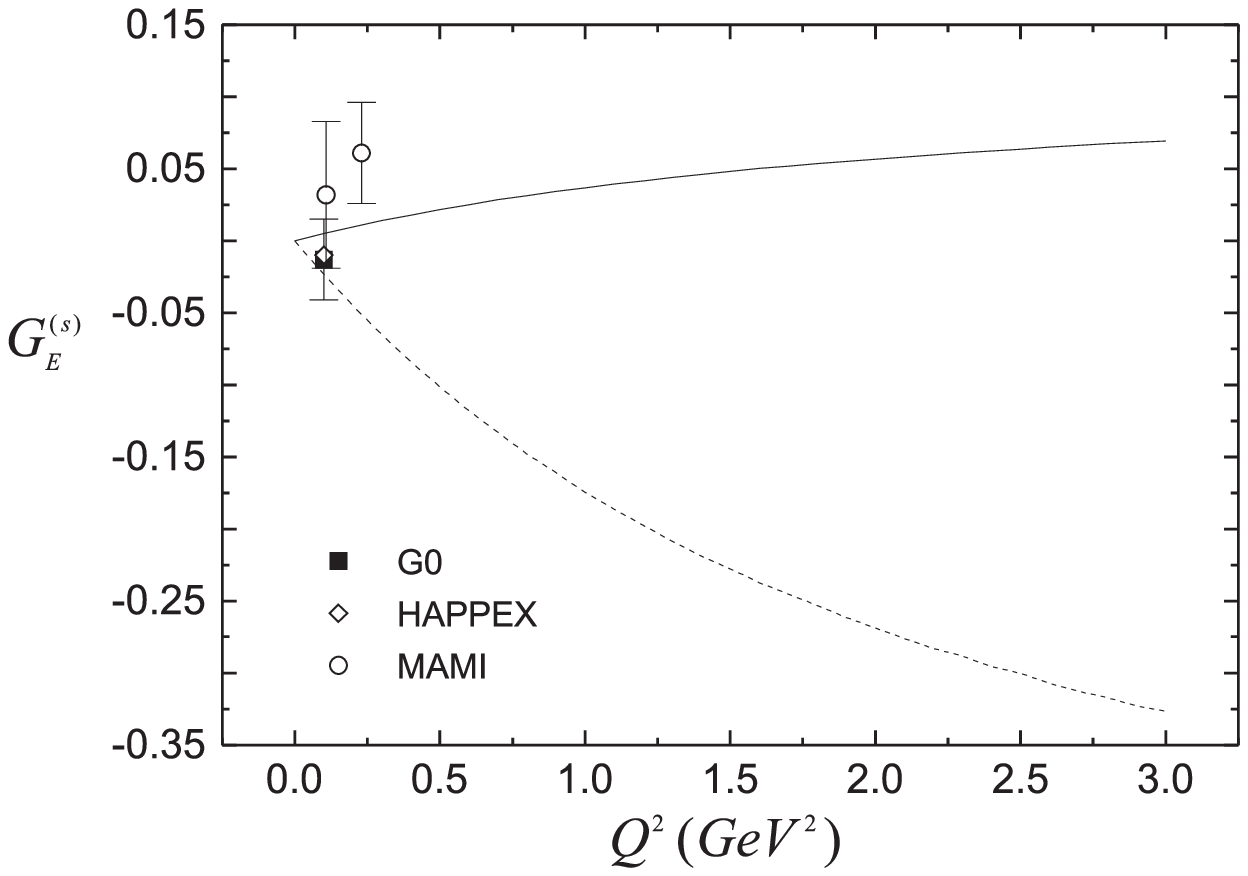,height=60mm,width=70mm,angle=0}
\psfig{figure=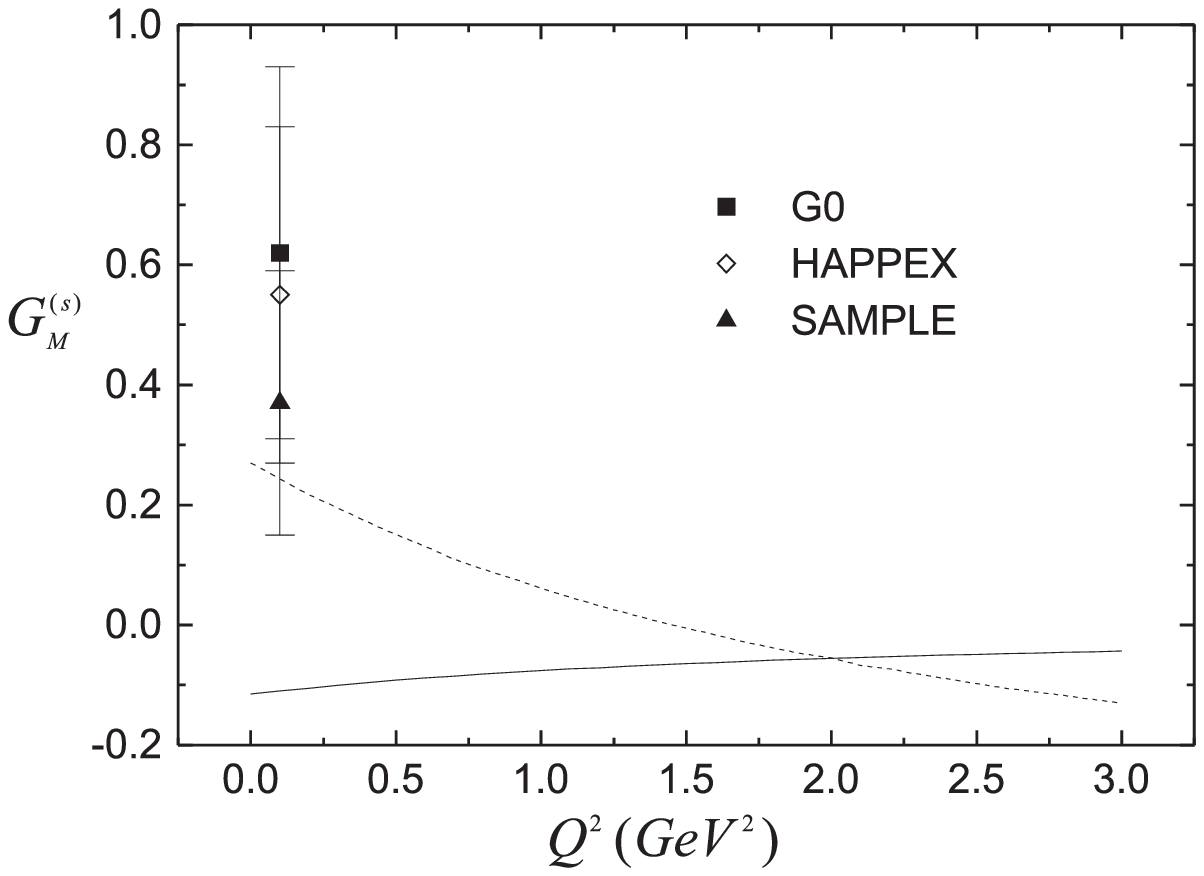,height=60mm,width=70mm,angle=0}}
\caption{The electric (left panel) and magnetic (right panel) strange form 
factors.
The solid and dashed lines show our  results with $\Lambda_K=\Lambda_{K^*}
=0.9\GeV$, $F_{KK^{\ast }}^{(s)}(0)=1.84$ and  $\Lambda_K=0.9\GeV,~\Lambda_{K^*}
=1.1\GeV$, $F_{KK^{\ast }}^{(s)}(0)=8.0$ respectively.} 
\end{figure}
Taking a closer look at the results obtained for the strange magnetic moment 
of the nucleon
obtained in ref.~\cite{hil00}, we see that while the contributions from the 
kaon and $ K^{\ast }$ are negative, the $K/K^{\ast }$ transition contribution 
is positive. Therefore, if one allows the $F_{KK^{\ast }}^{(s)}(0)$ coupling
in Eq.(\ref{spinfl}) to be bigger, it is possible to get a positive value
for $G_M^s(Q^2)$. Just to give an example, using $F_{KK^{\ast }}^{(s)
}(0)\sim8.0$, $\Lambda_{K^*}=1.1\GeV$ and keeping the other parameters fixed
we get at $Q^2\sim0.1~\GeV^2$: $G_E^s=-0.023$ and $G_M^s=0.24$. What is even
more interesting is the fact that the result for the combination $G_E^s+\eta~
 G_M^s$ remains almost unchanged, up to $Q^2\sim1~\GeV^2$, as compared with 
the dashed line in Fig.~1. This shows that it is very important to determine 
experimentally
each one of the strange for factors, and not only the combination
$G_E^s+\eta~G_M^s$, if one does really intend to understand the strangeness 
of the nucleon. In Fig. 2 we show, together with the available 
experimental data, our  results for the electric and magnetic
strange form factors using $\Lambda_K=\Lambda_{K^*}
=0.9\GeV$, $F_{KK^{\ast }}^{(s)}(0)=1.84$ (solid line) and $\Lambda_K=0.9\GeV,
~\Lambda_{K^*}=1.1\GeV$, $F_{KK^{\ast }}^{(s)}(0)=8.0$ (dashed line).

In summary, we have calculated the electric and magnetic strange form factors
of the nucleon with a version of the meson  cloud model, which includes the 
kaon and the $K^*$ 
contributions. In contrast to other situations, in the present case the $K^*$ 
contribution did not cancel the kaon contribution. Instead it reinforced it.
In our approach the combination in Eq.(\ref{comb}) is negative and decreasing
with $Q^2$. However, it is important
to point out that other version of the MCM, like the light-cone chiral cloud
model in ref.~\cite{mel99}, gives a positive value  to the combination in 
Eq.(\ref{comb}).

\vspace{.5cm}
 
\underline{Acknowledgments}: 
This work has been supported by CNPq and FAPESP (Brazil).


\end{document}